\documentstyle[11pt,aaspp4]{article}
\slugcomment{To appear in v540 of {\it The Astrophysical Journal} (09/01/00)} 
\lefthead{Wallace et al.}
\righthead{Variability of EGRET Sources}
\begin{document}

\title{A Systematic Search for Short-term Variability \\
       of \\
       EGRET Sources}

\author{P. M. Wallace, N. J. Griffis\altaffilmark{1}}
\affil{Department of Physics \& Astronomy, Berry College, Mt. Berry, GA 30149}

\author{D. L. Bertsch, R. C. Hartman, D. J. Thompson}
\affil{Code 661, NASA Goddard Space Flight Center, Greenbelt, MD 20771}

\author{D. A. Kniffen\altaffilmark{2}}
\affil{Department of Physics \& Astronomy, Hampden-Sydney College, Hampden-Sydney, VA 23943}

\and

\author{S. D. Bloom\altaffilmark{3}}
\affil{IPAC, JPL/Caltech, MS 100-22, Pasadena, CA 91125}

\altaffiltext{1}{present address: Dept. of Mechanical Engineering, Georgia 
                 Institute of Technology, Atlanta, GA 30332} 

\altaffiltext{2}{present address: NASA HQ, Washington, DC 20546} 

\altaffiltext{3}{present address: Dept. of Physics \& Astronomy, Hampden-Sydney College, Hampden-Sydney,
                 VA 23943} 

%

\begin{abstract}

The 3rd EGRET Catalog of High-energy Gamma-ray Sources contains 170 unidentified 
sources, and there is great interest in the nature of these sources. 
One means of determining source class is the study of flux variability on time 
scales of days; pulsars are believed to be stable on these time scales while 
blazars are known to be highly variable. In addition, previous work has 
demonstrated that 3EG J0241-6103 and 3EG J1837-0606 are candidates for a new gamma-ray 
source class. These sources near the Galactic plane display transient behavior 
but cannot be associated with any known blazars. Although many instances of 
flaring AGN have been reported, the EGRET database has not been
systematically searched for occurrences of short-timescale ($\sim1$ day) 
variability. These considerations have led us to conduct 
a systematic search for short-term variability in EGRET data, covering all 
viewing periods through proposal cycle 4. Six 3EG catalog sources are reported here
to display variability on short time scales; four of them are unidentified. 
In addition, three non-catalog variable sources are discussed. 

\end{abstract}


\section{Introduction}

The Energetic Gamma-Ray Experiment Telescope (EGRET) on board the {\it Compton 
Gamma-Ray Observatory} ({\it CGRO}) began observing the high-energy gamma-ray sky
in 1991. Since that time, EGRET has contributed enormously to the list of 
gamma-ray point sources; the third EGRET catalog (\cite{Har99}, hereafter H99) 
contains 271 point sources exhibiting significant flux above 100 MeV. Besides 
a single solar flare, the Large Magellanic Cloud (LMC), and a probable 
association with a radio galaxy (Cen A), the identified sources are 
distributed among two established classes of high-energy gamma-ray emitters: 
pulsars and radio-loud blazars. 170 sources remain unassociated with known 
objects. There is great interest in the nature of the unidentified sources, whether 
they be pulsars, blazars, or representatives of new classes of gamma-ray sources.  

The limited spatial resolution of EGRET often hampers source identification based 
solely on position; an alternate means of determining source class is the study 
of flux variability. In particular, pulsars are believed to be stable on short 
($\sim 1$ day) time scales (\cite{Ram95}) while blazars are known to be strongly variable 
(e.g., \cite{Mat97a}). Short-term and long-term variability studies have also led to the 
discovery of two candidates for a new gamma-ray source class: 3EG J1837-0606 (GRO J1838-04) 
(\cite{Tav97}) and 3EG J0241-6103 (2CG 135+01) (\cite{Kni97,Tav98}). These sources 
display transient behavior but cannot be associated with any known radio-loud 
blazars. 

There are many instances of gamma-ray flares of AGN on short time scales. 
For example, 3C 279 (\cite{Kni93,Weh98}), PKS 0528+134 (\cite{Hun93}), 3C 454.3 
(\cite{Har93}), PKS 1633+382 (\cite{Mat93}), PKS 1406-076 (\cite{Wag95}), PKS 
1622-297 (\cite {Mat97a}), and CTA 26 (\cite{Mat95,Hal97,Mat00a}) 
all show strong evidence of variability on time 
scales of 1-3 days. Many other AGN show variations on time scales of weeks to 
months (\cite{von95,McL96,Muk97}). Although these many instances of short-term 
variability have been noted, no exhaustive survey of the EGRET database has been performed. 

It is the purpose of this study to conduct a systematic and 
comprehensive survey of EGRET data from proposal cycles 1-4 in search of evidence of 
short-term flux variability above 100 MeV. The standard method of EGRET data analysis, 
maximum likelihood, was applied to short-duration (one-day and two-day) maps of gamma-ray 
intensity. The variability statistic $V$ was used as a guide for singling out those 
light curves which deviate appreciably from constant flux. For the unidentified 
sources that display evidence of flaring, Monte Carlo simulations were performed to 
determine the probability of such fluctuations assuming intrinsically nonvariable 
sources. Six catalog sources were found to exhibit variability on one- and two-day 
time scales: four are unidentified, one is a flat-spectrum radio quasars (FSRQ), 
and one is a BL Lac object. In addition, two transient non-catalog excesses and 
one non-catalog sub-threshold repeating source are discussed. The limitations of this 
study are then considered; a summary and concluding remarks follow.

\section{EGRET Data and Preliminary Analysis}

The data used for the present analysis are those from Phase 1
through Cycle 4 of the CRGO mission (1991 April - 1995 October),
matching the time span of the third EGRET catalog (H99). The CGRO timeline
is divided into observation (viewing) periods at fixed 
attitudes that range in duration from one day to several weeks. Each of the
144 viewing periods longer than 3 days was first divided into two-day 
intervals (for viewing periods with an odd number of days, the 
last interval has three days).  For each interval, a
sky map was generated for $E>100$ MeV using the same criteria used to
generate standard EGRET maps.  In addition, for each instrument pointing
within 20$^{\circ}$ of the Galactic plane (where the sources typically are
brighter), a series of one-day maps was generated. Each of these maps was 
analyzed with the EGRET maximum likelihood program (\cite{Mat96}) 
using the following procedure: (1) All sources from H99 within 30$^{\circ}$ 
of the pointing direction were modeled.  Sources beyond 30$^{\circ}$ for 
these short intervals have too little exposure to be useful and would 
require separate analysis, because the EGRET Point Spread Function (PSF) 
is broader at wide angles (\cite{Esp99}). (2) Fluxes for all the sources 
and the EGRET diffuse radiation model (\cite{Hun97,Sre98}) within 
the map were optimized simultaneously, producing a table with statistical 
significance and flux for each source. (3) A search was made for any 
statistically-significant source in the map not found in the EGRET catalog.

The statistics of the data are very limiting, especially for the one-day 
light curves. In order to isolate the curves for which the flux uncertainty 
is not overwhelming, the significance of each of the short-duration detections 
were examined. An excess of 4$\sigma$ is the minimum requirement for inclusion 
of a source in H99, and this threshold was applied here so that 
only those light curves which contain at least one 4$\sigma$ short-duration 
detection were considered for further analysis. 

Even with this requirement, much of the data suffers from large statistical 
uncertainty. This leads us to the conclusion that for all but the strongest 
sources, EGRET is sensitive to only the most dramatic short-term 
variations in flux. In order to single out the most variable 
light curves, the variability statistic $V$ (\cite{McL96}) was used. For the 
remaining light curves the $\chi^2$ statistic was calculated from

\begin{equation}
\chi^2 = \sum_{i=1}^{N}\frac{(F_i-\bar{F})^2}{\sigma_i^2}
\end{equation}

where N is the number of observations, $F_i$ is the detected flux during the $i$th 
observation, $\bar{F}$ is the mean flux for the viewing period, and $\sigma_i$ is
the 1-$\sigma$ flux uncertainty of the $i$th observation. If $Q$ 
is the probability of obtaining a value of $\chi^2$ equal to or greater than
the empirical $\chi^2$ from an intrinsically nonvariable source, then 
$V\equiv-\mathrm{log}$$Q$. For this work, light curves with $V\ge3.0$ ($Q\le10^{-3}$) 
are considered to manifest source variability. Light curves displaying one 
or more consecutive strong two-day detections or two or more consecutive strong 
one-day detections against a background of weak detections and/or upper limits 
are considered to show evidence of flaring. In such situations, the value of 
$V$ is depressed by the high number of weak detections and/or upper limits; for 
this reason these curves were retained for further study if they had variability 
indexes at or above 1.0.

Seven light curves of six sources displayed one or two-day 
variability indexes greater than 3.0 and/or 1.0 with evidence of flaring. 
With one exception, the one-day analysis turned up a final list of variable 
light curves identical to that of the (independently performed) two-day analysis. 
Four of the seven sources are listed as unidentified in H99.

For the unidentified sources that display evidence of flaring behavior, Monte Carlo 
simulations of the data were performed. The data in question were simulated using the 
current models of the diffuse isotropic radiation (\cite{Sre98}) and the diffuse 
Galactic emission (\cite{Hun97}) in conjunction with the source flux. In some cases, 
the source in question was not isolated and a number of nearby sources 
were included in the simulation; all sources were assumed to be intrinsically 
nonvariable. Mean fluxes for each source and short-duration 
exposures representative of the viewing period are input into the model. 
For a given source, the simulation consisted of distributing the flux as the EGRET PSF
and adding Poisson-distributed deviates pixel-by-pixel. 
In this way 1000 counts maps were generated; these maps were 
analyzed using maximum likelihood, resulting in a distribution of the significance for 
the source of interest. The probability of finding a single detection and subsequently 
a ``flare'' of given overall significance or greater from an intrinsically 
nonvariable source was then determined from this distribution. See Mattox et al. (1996) 
for more details about the Monte Carlo simulation.

\section{Discussion of Individual Catalog Sources}

The following sources are listed in H99 and display evidence for variable output. 
To be listed in H99, sources with $|b|<10^\circ$ must be detected with a 
significance of at least $5\sigma$ and sources with $|b|>10^\circ$ must be detected 
at a significance of at least $4\sigma$. Four of the sources have not been previously 
known to display short-term variability. The variability of the other two has been 
reported and for these we present additional work.

\subsection{New Reports of Short-term Variability}

\subsubsection{3EG J0222+4253 (3C 66A/PSR J0218+4232?)}

Source 3EG J0222+4253 exhibits a strong two-day flare in VP 15.0
\footnote{The preliminary version of this result appeared in Bloom et
al. 1997.}; virtually all of the detection in this 14-day interval came
on days 11 and 12.  The peak flux value on day 12, $(8.1\pm2.4)\times10^-7$
photons cm$^{-2}$ s$^{-1}$, is over four times the mean flux for this source,
either in the full viewing period or averaged over all observations in
H99. The source was 12$^{\circ}$5 off-axis and the two-day light curve 
(see Figure 1) registers a variability index of 2.6. Monte Carlo simulation of this 
source indicates that one should expect 
only about 5 in $10^4$ such viewing periods to display an equivalent or stronger 
``flare'' if the source is intrinsically nonvariable. Source 3EG J0222+4253 was also 
observed within 20$^\circ$ of the instrument axis in VP's 211.0, 325.0, and 427.0, 
but was found in all three cases to be nonvariable and consistent with the lower-level
flux observed in VP 15.0.

At energies above 1 GeV, the spatial analysis (H99) strongly supports
an association of this source with the BL Lac object 3C66A ($z=0.444$). 
This identification would place this source among the more than
60 blazar-class objects seen by EGRET, many of which show short-term
time variability, as noted above. 

Verbunt et al. (1996) and Hermsen et al. (1999) suggest that much of
the sub-GeV gamma-ray flux may have its origin in a second source, the
binary millisecond pulsar PSR J0218+4232, located less than 1 degree
from 3C66A. Their argument is based on three lines of evidence: (1) the
100-1000 MeV data show a 3.5 sigma evidence of pulsation at the radio
period; (2) the gamma-ray light curve resembles the one seen in hard
X-rays; and (3) the spatial analysis shows that the source position
moves from the BL Lac position toward the pulsar position with
decreasing gamma-ray energy. 

For the two-day flare, the EGRET data do not have good enough
statistics or resolution to determine which of the two candidate sources
is responsible for the flare.  Modeling both the AGN and the pulsar at
their known positions, the likelihood analysis assigns some photons to
each source.  The flare does have many more photons at lower energies
than higher energies, as indicated in Table 2. Also listed in Table 2 are
the number of photons expected from a single 2-day interval given the
average flux of 3EG J0222+4253 in VP 15.0. From the numbers it is 
clear that the flare is seen at all energies, including the 100-300 and 
300-1000 MeV bands. The energy distribution 
of the flare can also be seen in the source spectrum for all
of VP 15.0, shown in Figure 2.  The photon number index is $2.37\pm0.29$
for this viewing period, compared to $2.01\pm0.14$ for the sum of phases
1-4 for this source.  Because most of the photons that make up the flare
are of lower energy, they would be associated with the pulsar and not
the AGN in the two-source model.  Flaring behavior is not seen,
however,  in the other EGRET-detected pulsars.  We cannot rule out the
possibility that the AGN spectrum extended to lower energies during the
flare, although other EGRET AGN spectra tend to flatten rather than
steepen during flares (\cite{Sre96,Har00}).

\subsubsection{3EG J1410-6147}

In the first four days of VP 14.0, this unidentified source decayed from a 
flux of almost $(5.4\pm1.5)\times10^{-6}$ photons cm$^{-2}$ s$^{-1}$ to below 
EGRET's sensitivity, where it remained for the rest of the 14-day period
(see Figure 3). This is suggestive of flaring behavior. The one-day 
variability index is 1.5 and Monte Carlo simulation gives a probability of
about 0.0006 that the fluctuation found in VP 14.0 is produced by a 
nonvariable source. Although the source is 27$^\circ$1 off-axis 
during VP 14.0, this was an early observation, when EGRET's sensitivity was high.

It has been suggested that 3EG J1410-6147 ($l=312.18$, $b=-0.35$) is 
associated with SNR G312.4-0.4 (\cite{Stu95}), which falls just outside the
68\% error contour; this association is not well-confirmed. Flaring behavior 
would not be expected if the origin of the gamma rays was either cosmic rays 
from the SNR interacting with the surrounding material or a radio-quiet pulsar 
in the SNR. No association with any point radio source can be made (\cite{Mat00b}).
Sources 3EG J0241-6103 (2CG 135+01; see below) and 3EG J1837-0606 
(GRO J1838-04) share this characteristic (\cite{Kni97,Tav98}) of low Galactic latitude,
variability, and 
lack of a strong radio counterpart. It may be that along with these sources, 
3EG J1410-6147 represents a new class of gamma-ray 
emitter. This suggestion should be considered with caution in light of the high 
aspect during VP 14.0.

\subsubsection{3EG J1746-2851}

As this source is unidentified, strong, and coincident with the Galactic 
Center, it has been studied in some detail (\cite{May98}). However, until now 
its short-term time history has not been examined. There is some evidence 
of fluctuation in VP's 16.0 and 429.0. It should be stated first that 3EG 
J1746-2851 sits in the center of the most densely-packed region of the gamma-ray sky; 
there are 10 sources listed in H99 within 10$^\circ$ of the Galactic Center. 
Given the broad EGRET PSF, source confusion is a serious problem. In 
addition, EGRET's sensitivity was high during this early observation but 
the source was 20$^\circ$3 off-axis, leading to increased statistical 
uncertainty. However, while 3EG J1746-2851 appears to fluctuate during two 
different viewing periods, no other sources in highly congested regions display any 
evidence of short-term variability.

In VP 16.0
the three strongest one-day detections fall on days 7-9 of the 14-day period, 
exhibiting a peak flux of $(3.40\pm0.92)\times 10^{-6}$ photons cm$^{-2}$ 
s$^{-1}$ on day 8 (see Figure 4). This is only a 4.3$\sigma$ detection, but it 
is flanked on days 7 and 9 by detections of 3.9$\sigma$ and 3.1$\sigma$ 
respectively and there is no detection stronger than 2.3$\sigma$ within 5 days 
of the peak. The one-day variability index for VP 16.0 is 2.1, corresponding to 
a probability of 0.008 that these data are consistent with a nonvariable source. 
Monte Carlo analysis is more restrictive. This source and the seven others within 
a 7$^\circ$ radius were modeled and there is a probability of 0.0011 that a 
fluctuation of this (or greater) significance will occur in a 14-day 
period given intrinsically nonvariable sources. 

Source 3EG J1746-2851 also shows evidence of variability in VP 429.0. During this 
pointing the aspect is only 6$^\circ$ and the one-day variability index has 
a value of 3.0. The peak flux is very high $(6.4\pm1.7)\times10^{-6}$ 
photons cm$^{-2}$ s$^{-1}$ and on two days the source is not detected at all, 
but there is no evidence of flaring. The light curve is shown in Figure 5.

Time variability would indicate a compact object as the gamma-ray
source.  Models such as advection-dominated accretion flow onto a
massive black hole (\cite{Mah97}) would be favored rather than a
collection of pulsars or a cosmic ray enhancement.  See
Mayer-Hasselwander et al. (1998) for further discussion of possible
models.

\subsubsection{3EG J2006-2321}

During VP 13.1 (1991 October 31 - November 7), this source ($l=18^\circ8, 
b=-26^\circ3$) exhibited transient behavior on time scales of 12 hours. During 
this VP the source was 13$^\circ$6 from the instrument axis and the overall significance 
of the detection is 4.8$\sigma$. A light curve of 3EG J2006-2321 in VP 13.1 
is shown in Figure 6. The first two days of observation are broken down 
into four 12-hour periods; the remaining points represent full days. Breakdown 
of the data into smaller time intervals is not useful as poor statistics 
become overwhelming. EGRET appears to have caught 3EG J2006-2321 on the leading 
edge of an intense flare, although this is not known with high confidence. 
The peak flux, centered on MJD 48560.25, is $(1.75\pm0.53)\times10^{-6}$ 
photons cm$^{-2}$ s$^{-1}$; this 12-hour detection has a significance of 5.4$\sigma$. 
The source then decays monotonically on this time scale; less than two days later 
it is not detected by EGRET at all. The flux never rises above 
$3.9\times10^{-7}$ photons cm$^{-2}$ s$^{-1}$ for the rest of VP 13.1. 
The ratio of peak to average flux for the VP is 5:1. The gamma-ray spectrum of 
3EG J2006-2321 for VP 13.1 is consistent with a power law with exponent 
$-2.13\pm0.31$. 

Applying a $\chi^2$ test to the light curve yielded a probability of 0.0006 
that these data are consistent with an intrinsically nonvariable source. 
Results from Monte Carlo simulations of the source are consistent with these analyses. 
From the simulated data it was determined that the probability of finding such a 
deviation in a 7-day viewing period is 0.0005.

The combination of high Galactic latitude and short-term transient behavior 
suggests an association with the blazar class of AGN. While this source may well 
be a blazar, its radio flux at 5 GHz is significantly lower than blazars with similar
peak gamma-ray flux values. The best candidate for association with 3EG J2006-2321 
is the radio source PMN J2005-2310 (260 mJy at 5 GHz, $\alpha_r$ not known at this 
frequency; (\cite{Gri93})). Of the 10 high-confidence blazars listed by Mattox et al. 
(1997) with peak gamma-ray flux above $10^-6$ photons cm$^{-2}$ s$^{-1}$, all are 
associated with flat-spectrum radio sources having a 5 GHz intensity above 1.0 Jy. 
The analysis of Mattox et al. (2000b) has determined that the {\it a priori} probability 
of EGRET detecting such a weak radio source is 0.0006. This low probability ensures that even 
though PMN J2005-2310 falls only 11.0 arcmin from the EGRET-determined position of 3EG 
J2006-2321, well within the 50\% confidence countour, the probability of association 
is only 0.015 (\cite{Mat00b}). However, the variability presented in the present study bolsters the
possibility of the source being a blazar. In addition, while PMN J2005-2310 is not bright, 
there is strong evidence that it is a flat-spectrum source; it is listed with a flux of 
260 mJy in the Texas 365 MHz survey (\cite{Dou96}) where it is listed as TXS 2002-233. 
The spectral index around 365 MHz is $0.7\pm0.2$. The radio source also appears as NVSS 
J200556-231028 in the NRAO VLA Sky Survey (\cite{Con98}) at 1.4 GHz with a strength of 302 mJy.

A multi-wavelength investigation of PMN J2005-2310 is encouraged in order to determine 
its nature. VLBI measurements may resolve a compact core, and optical observations 
may aid in determining redshift and polarization. Finally, an X-ray detection of 
this source will help to establish its spectral energy distribution.

\subsection{Follow-up Reports of Short-term Variability}

\subsubsection{3EG J0241+6103 (2CG 135+01)}

Source 3EG J0241+6103 is unidentified and has been the subject of several 
papers because of evidence of short-term variability and lack of 
radio-loud blazar counterparts (\cite{Kni97,Tav98}). The short-term data have 
been re-analyzed in light of indications that another source (3EG J0229+6151, 
H99) exists only 1$^\circ$68 away with about half the intensity of 
3EG J0241+6103. The results still indicate evidence of variability in VP 211.0 
(1993 25 February - 9 March) with a two-day variability index of 3.1 (see Figure 7). 
Although this source is near the galactic plane ($b=+0^\circ$99), there are no 
EGRET-detected sources other than 3EG J0229+6151 within $10^\circ$. Source 3EG J0241+6103   
and the nearby source were modeled using Monte Carlo methods and the probability 
of finding a fluctuation this strong or stronger from steady sources 
in a 14-day period is about $5.9\times10^{-4}$. During no viewing period other than 
211.0 was the source found to be variable.

\subsubsection{3EG J0530+1323 (PKS 0528+134)}

Previous analyses of this quasar, located near the Galactic anticenter, found 
it to be strongly variable on two-day 
time scales during early EGRET observations (\cite{Hun93,Tho97}). PKS 0528+134
was observed in 1993 March to have a peak flux of $3.5\times10^{-6}$ photons cm$^{-2}$
s$^{-1}$; by 1993 May it had decreased by a factor of 2.5. The two-day 
variability index of 4.6 for VP's 0.2 through 2.1 corresponds to a probability of 
0.0002 that these data are consistent with an intrinsically nonvariable 
source. This is in stark contrast to the results of almost three weeks of continuous 
observations of the Galactic Anticenter in 1995 February - March (VP's 412.0 and 
413.0). During this time PKS 0528+134 displayed virtually constant flux 
in its two-day light curve (see Figure 8); there is no evidence of variability from 
this strong, previously erratic source. The $\chi^2$ analysis indicates that the 
probability that these late-period data are consistent with a constant source is 0.97. 
Therefore one must be careful when using flux histories to determine source 
class; while strongly variable data might be used to rule out identification of 
pulsars, measurements of constant flux do not rule out identification of AGN. 

\section{Discussion of Non-catalog Sources}

The thresholds for inclusion in H99 ($5\sigma$ for $|b|<10^\circ$ and $4\sigma$ 
for $|b|>10^\circ$) apply to any single viewing period or combination of viewing periods; 
however, it is 
possible for a source to display a short-duration significance well above that 
of the entire viewing period. This is almost always true of flaring sources, and is true of 
the following two sources. Their overall detections during the viewing periods in question 
fall below the catalog thresholds, but have short-duration excesses above these 
values.

\subsection{GRO J0927-41}

This source appears near $l=266.68$, $b=7.03$ with a 95\% confidence radius of 
1$^\circ$01 during the third and fourth days of VP 338.5. On these days the 
source displays excesses of $3.3\sigma$ and $4.0\sigma$ respectively (see Figure 11). 
The combined significance for these two days is $5.2\sigma$. The very strong Vela 
pulsar is 10$^\circ$4 away; Monte Carlo simulation of GRO J0927-41 and Vela
indicates that the probability of such transient behavior from
an intrinsically nonvariable source is 0.0013. A search for counterparts revealed
PKS 0920-397, a FSRQ 1$^\circ$33 away near the 99\% confidence contour. 
This quasar ($z=0.591$) has a flat radio spectrum above 1.4 GHz, with a flux of 
1.5 Jy at 4.85 GHz. It may be that EGRET has detected the tip of a flare from 
this source. There are not enough counts in the data to produce a spectrum of this 
source. 

\subsection{GRO J1547-39}

In VP 226.0, an excess of $4.7\sigma$ appears near $l=336.17$, $b=11.78$ 95\% 
with a confidence radius of 0$^\circ$97. The source is given the name GRO J1547-39 and
is seen most clearly in the third of four two-day maps. The light curve is
shown in Figure 10. When broken down into one-day intervals, the 
significance falls to $4.0\sigma$ and $3.1\sigma$. The overall detection for 
this source in VP 226.0 is $3.6\sigma$; in no other viewing periods that include the source 
position does the overall significance rise above $3.0\sigma$. No flat-spectrum
radio loud blazars or other candidate objects appeared as candidates for 
association with this source, and there are not enough counts to provide 
a spectrum. Monte Carlo simulation of this source indicates that there is a probability of 
0.010 of finding a ``flare'' of equal or greater strength from an intrinsically 
nonvariable source. With such a high probability we make only a weak claim for variability 
of GRO J1547-39.

\subsection{Sub-threshold Repeating Sources}

In the region $-40<b<40$, 148 two-day non-catalog excesses with 
significance $\geq3.5\sigma$ are found. Among these are seven spatially 
coincident pairs and one spatially coincident triplet. For 
each set, the data are combined; in six of these combined maps the 
significance exceeds the standard thresholds. These pairs are 
considered for further analysis, with the hypothesis that each one 
represents the same transient source at high flux values.

This hypothesis is almost certainly not true. The probability of finding 
six such coincident pairs from 148 excesses is not negligible; assuming 
that the sources are randomly scattered and that each source has an error 
circle with radius $\sim$1$^\circ$0, the probability of finding six or more 
spatially coincident pairs is nearly 0.03. Beyond this simple analysis, 
the individual excesses are below threshold and most are close to the 
Galactic plane, and may not represent real sources. However, of these 
six pairs one is worthy of mention for its high combined significance. 

%


During VP 423.5, a source appeared near $l=334.78$, $b=-0.28$, prompting a
report to the IAU (\cite{Kan95}). The source, GRO J1627-49, has a 95\% 
confidence radius of 0$^\circ$42 and was found to be coincident with an H \small II
\normalsize region. However, it was not seen in any other band and faded during the 
viewing period; it was not strong enough for the viewing period as a whole to be 
included in H99. The present analyses reveal that during VP 23.0 a sub-threshold 
excess appeared for two days at a coincident position. The data from these viewing 
periods combine to yield an excess with significance $5.7\sigma$. The spectrum of 
these combined data is shown in Figure 11 and is rather flat, with a spectral 
index of $1.78\pm0.31$. 

\section{Limitations of this Study}

EGRET's sensitivity, while higher than any previous gamma-ray telescope, is low 
enough to severely limit the present analyses when combined with the intrinsically 
low gamma-ray flux. Often there are simply not enough counts to break t
he data down into smaller divisions 
and retain any meaningful information. For much of the data, low counts translate 
into statistical errors of $\sim$40-50\%. In these cases, an intrinsic flux 
increase of 100\% is barely detected with any significance; changes of 
$\sim$200-300\% are required for higher-confidence claims of variability. 
Thus only the most dramatic changes in flux are measured with any confidence. 
For all but the strongest sources, the estimated 10\% systematic uncertainty 
(\cite{Tho95}) adds insignificantly to statistical error. 

Interpretations of the variability index $V$ are rendered problematic by 
the low EGRET count rate. A low value of $V$ may have its origin in 
constant source output; it may also be due to poor statistics. There is 
no way to distinguish between these two cases. Again, this means that 
only very large values of $V$ are meaningful, and that only strong variations 
from constant flux are detected.

Also, the high short-duration errors lead to uncertainties in mean fluxes 
for given viewing periods; the shorter the duration of the detection, the higher the 
uncertainty of the mean. This may result not only in misleading values of $V$ 
but also Monte Carlo probabilities, as these calculations assume the averages 
for the viewing period in question. 

\section{Summary and Conclusions}

The results of this work are summarized in Table 1. The first column indicates 
the source name by increasing Right Ascension; sources found in H99 are given the 3EG 
designation and non-catalog sources are given the GRO designation.
The second column indicates alternate names, and the 
third column indicates source identification as FSRQ, BL Lac, 
or unidentified (UID). Column four indicates those viewing periods in which 
the source is found to be variable, column five gives the variability index for 
that viewing period, and column six indicates whether the source flared in that 
viewing period. The Monte Carlo probability for unidentified flaring sources is given 
in the seventh column.

The detections of short flares in 3EG J0222+4253 and 3EG J2006-2321 demonstrate that
EGRET can detect short, intense flares in relatively weak catalog sources 
($3.9\sigma$ and $4.8\sigma$ in the viewing periods in which the flares were seen) as well
as in strong blazars. Therefore most EGRET sources, except bright AGN, are not 
characterized by strong variability on one- and two-day time scales. If intense flares from 
sources like 3EG J0222+4253 and 3EG J2006-2321 were common, this survey would have detected 
more of them.

The same conclusion is reached for sub-threshold sources. The detections of 
GRO J0927-41 and GRO J1547-39, plus the repeating sources, demonstrate 
that EGRET can find very weak transients from sub-threshold sources with this 
sort of systematic analysis. The fact that few such intense flares are seen in this 
study indicates that the gamma-ray sky is not filled with these kind of flares 
from sub-threshold sources.

The flaring behavior seen in 3EG J0222+4253 adds another puzzle to
the issue of identification of this source.  The flaring is seen at
energies between 100 MeV and 1 GeV, where the emission may be dominated
by PSR J0218+4232. This would make this source unusual in two ways; it would be 
the only ms pulsar seen by EGRET and the only pulsar to show strong flaring.

The long observation of steady emission from PKS 0528+134 shows that, as seen by
EGRET, blazars are not always variable even on time scales of three weeks. 
Therefore long episodes, free of intense flares from a given source,  
do not rule out the identification of that source as a blazar.

The observation of clear short-term variability in 3EG J2006-2321 indicates that
this source may be a blazar, although its radio flux at 5 GHz is significantly less
than other EGRET-detected blazars with similar peak gamma-ray brightness.
The suggestion of variability of 3EG J1410-6147 may place it alongside 
3EG J0241-6103 and 3EG J1837-0606 as a third candidate for a new gamma-ray class
of variable Galactic sources.

In summary, the results of this study suggest that EGRET reached the
threshold of an important new feature of gamma-ray astrophysics: the
ability to measure short-term time variability.  The few examples found
by EGRET show that such sources do exist and are not restricted to the
blazar class.  Variability offers an effective tool for
matching gamma-ray sources to objects seen at other wavelengths.  With
its ability to detect much fainter sources, measure variability on
time scales of hours, and detect smaller flux variations that EGRET, 
the proposed Gamma-ray Large Area Space Telescope (GLAST) should be able 
to take full advantage of gamma-ray source time variability to extend the 
results of the present study.

\acknowledgments

The authors wish to thank J. R. Mattox and J. P. Halpern for their helpful comments 
and suggestions. P. W. gratefully acknowledges support from the NASA/ASEE 
Summer Faculty Fellowship Program.

\begin{figure}[t]
\epsscale{0.85}
\plotone{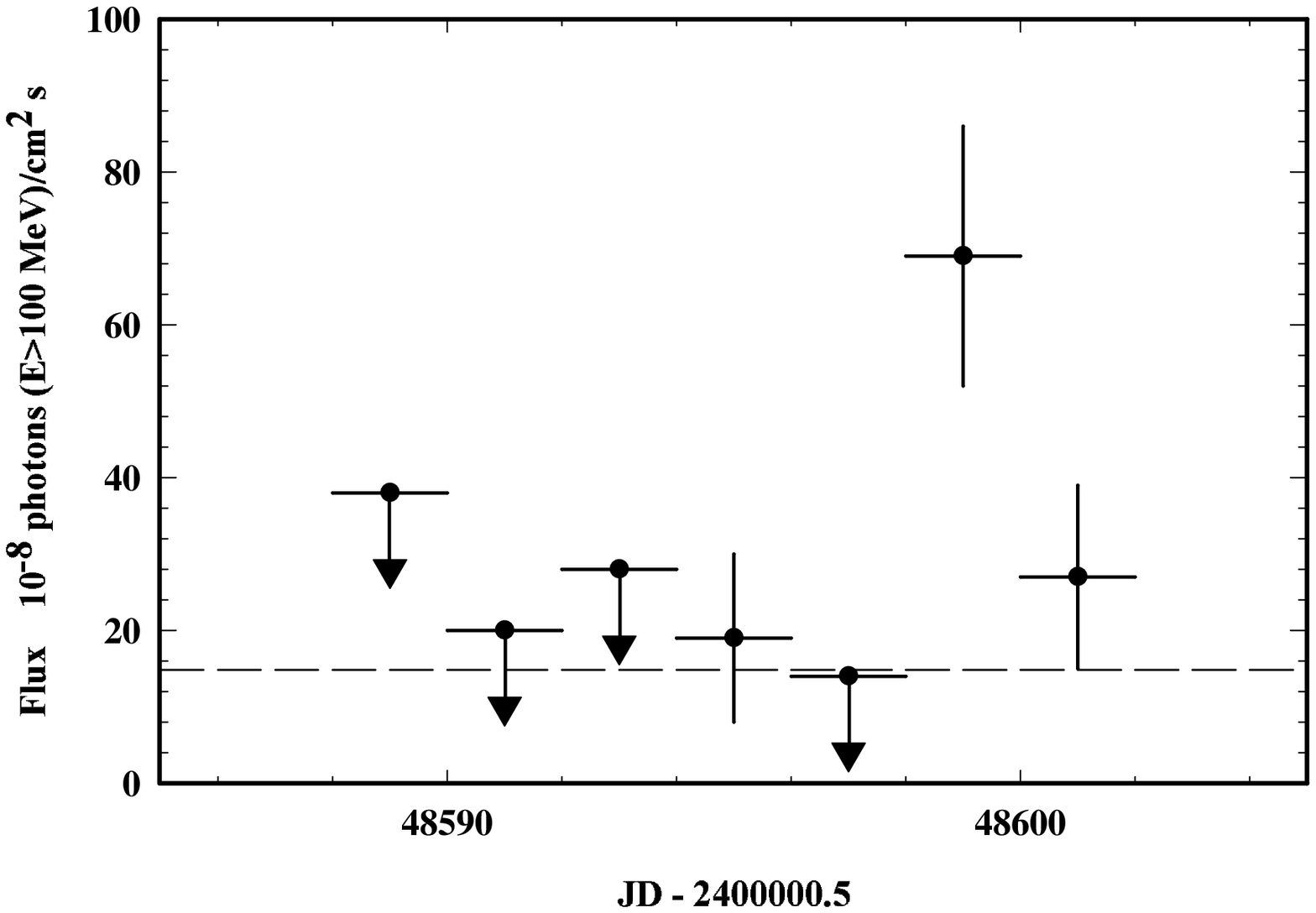}
\caption{Light curve of 3EG J0222+4253 (3C 66A) during VP 15.0. The points represent
         48-hour integration times. The variability index $V = 2.6$.\label{fig1}} For
         all light curves, the dashed line indicates the average flux for the viewing 
         period.
\end{figure}
\clearpage

\begin{figure}
\epsscale{0.6}
\plotone{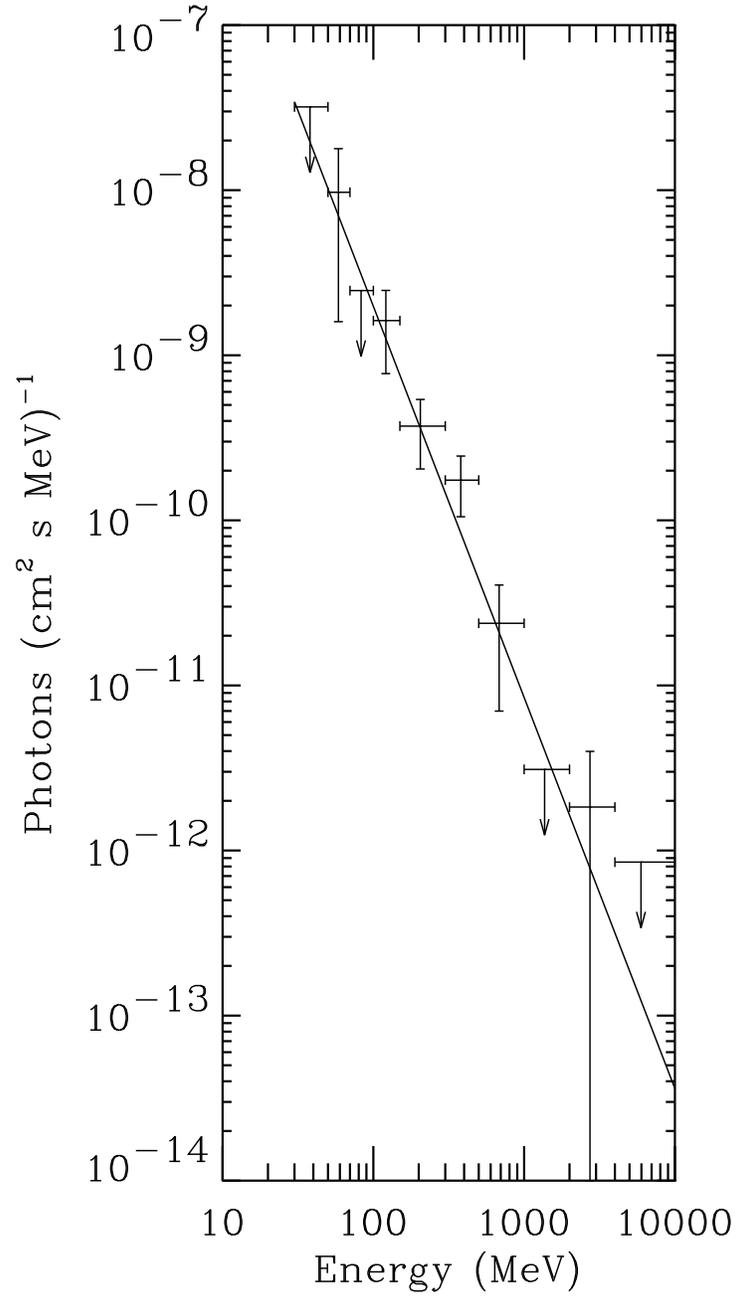}
\caption{Spectrum of 3EG J0222+4253 (3C 66A) during VP 15.0. The spectral index is 
         $2.37\pm0.29$.\label{fig2}}
\end{figure}

\begin{figure}[t]
\epsscale{0.85}
\plotone{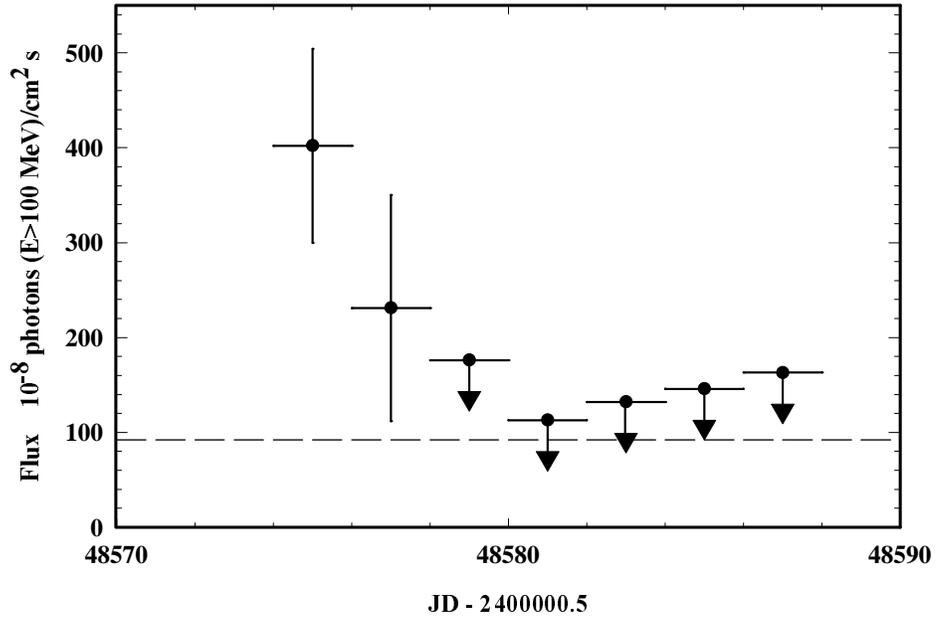}
\caption{Light curve of 3EG J1410-6147 during VP 14.0. The points represent
         48-hour integration times. The variability index $V = 2.4$.\label{fig3}}
\end{figure}
\clearpage

\begin{figure}[t]
\epsscale{0.85}
\plotone{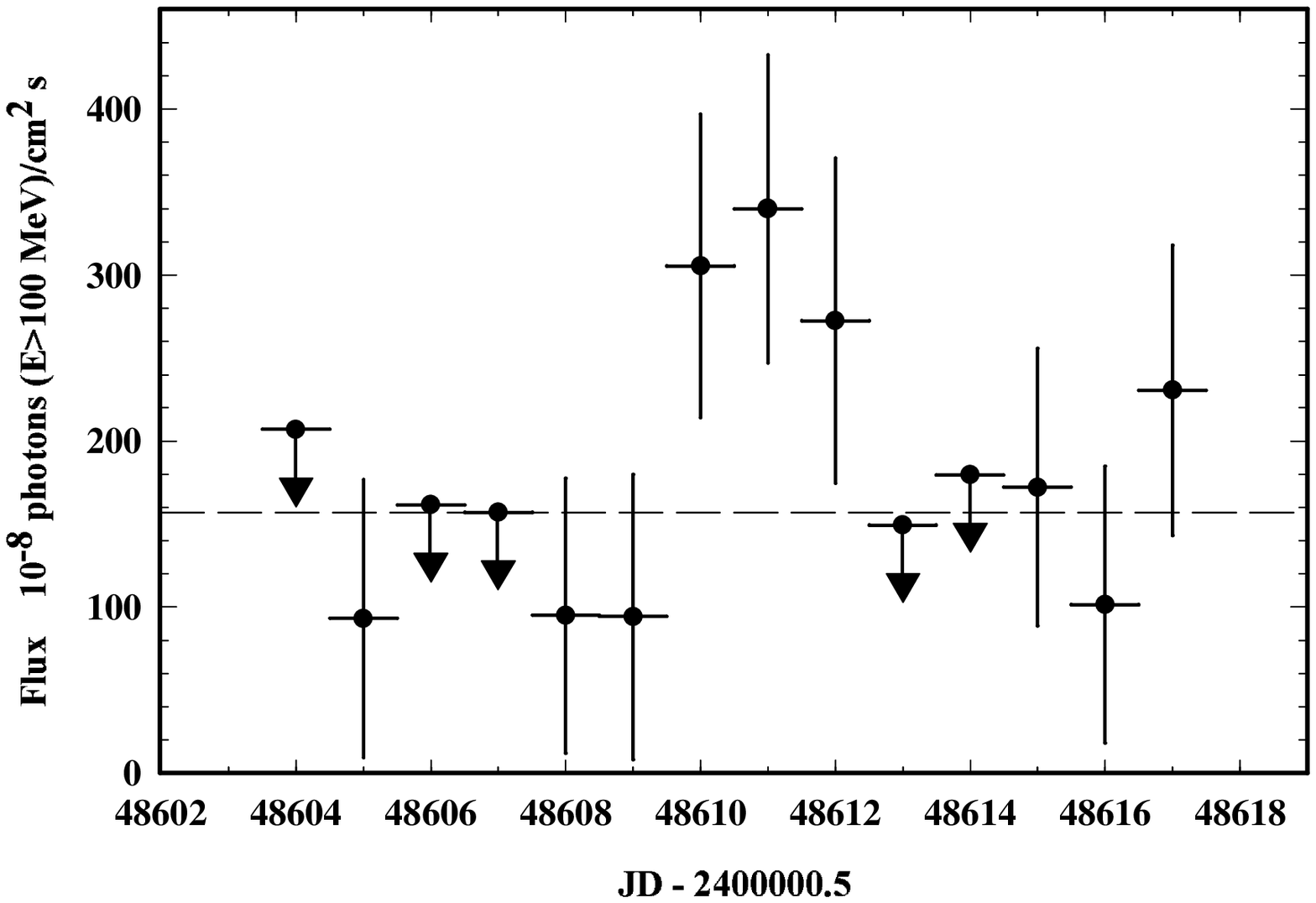}
\caption{Light curve of 3EG J1746-2851 during VP 16.0. The points represent
         24-hour integration times. The variability index $V = 2.1$.\label{fig4}}
\end{figure}
\clearpage

\begin{figure}[t]
\epsscale{0.85}
\plotone{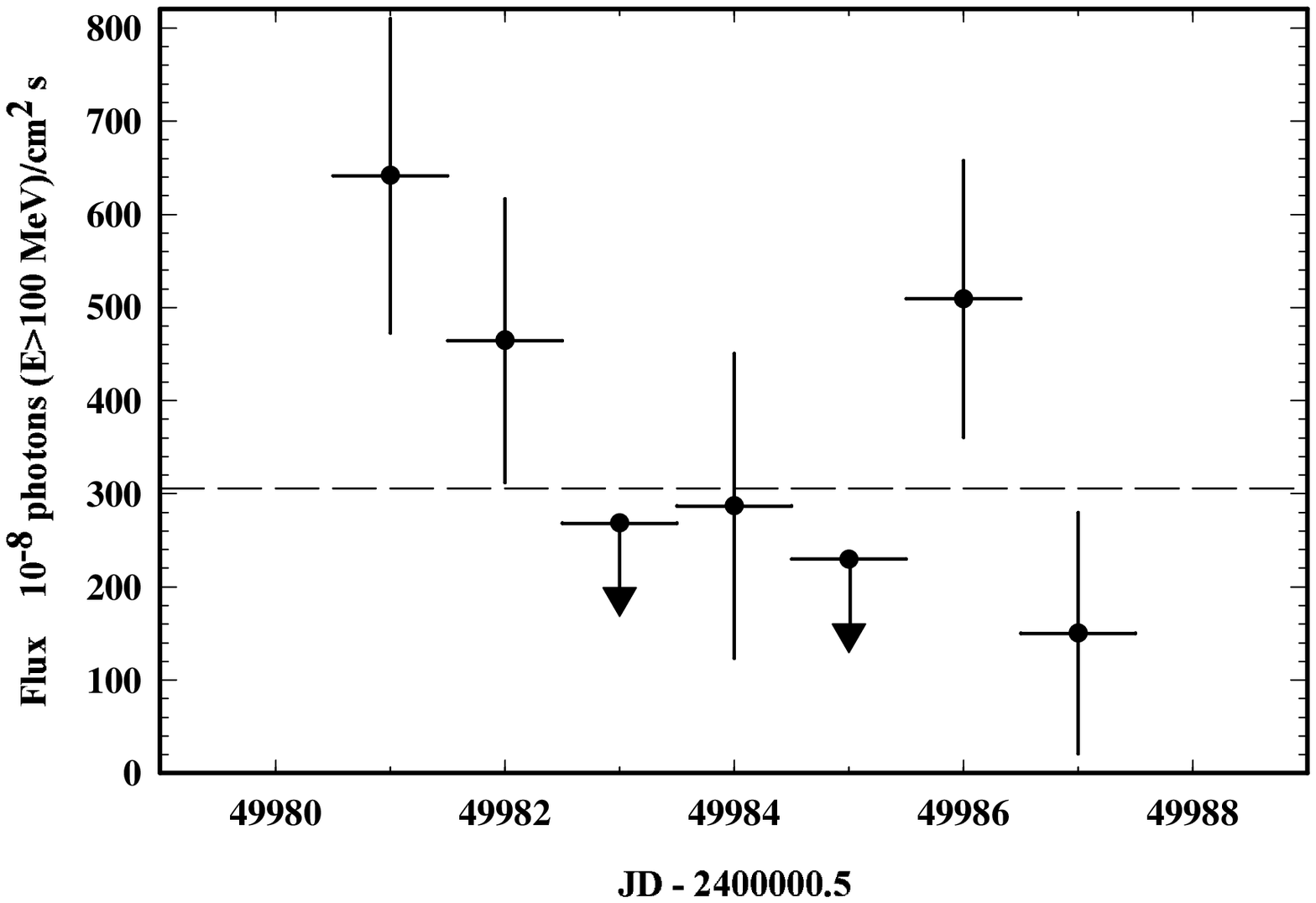}
\caption{Light curve of 3EG J1746-2851 during VP 429.0. The points represent
         24-hour integration times. The variability index $V = 3.0$.\label{fig5}}
\end{figure}
\clearpage

\begin{figure}[t]
\epsscale{1.0}
\plotone{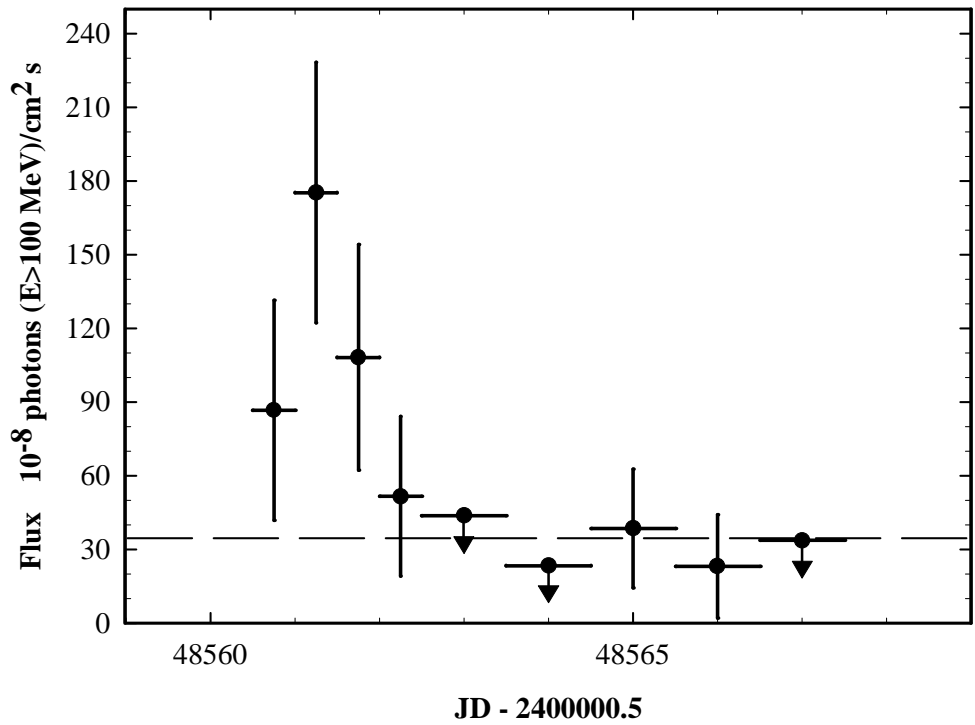}
\caption{Light curve of 3EG J2006-2321 during VP 13.1. The first four points represent
         12-hour integration times and the last five points represent 24-hour integration
         times. The dashed line indicates the average flux for the viewing period. The 
         variability index $V = 3.2$.\label{fig6}}
\end{figure}

\begin{figure}[t]
\epsscale{0.85}
\plotone{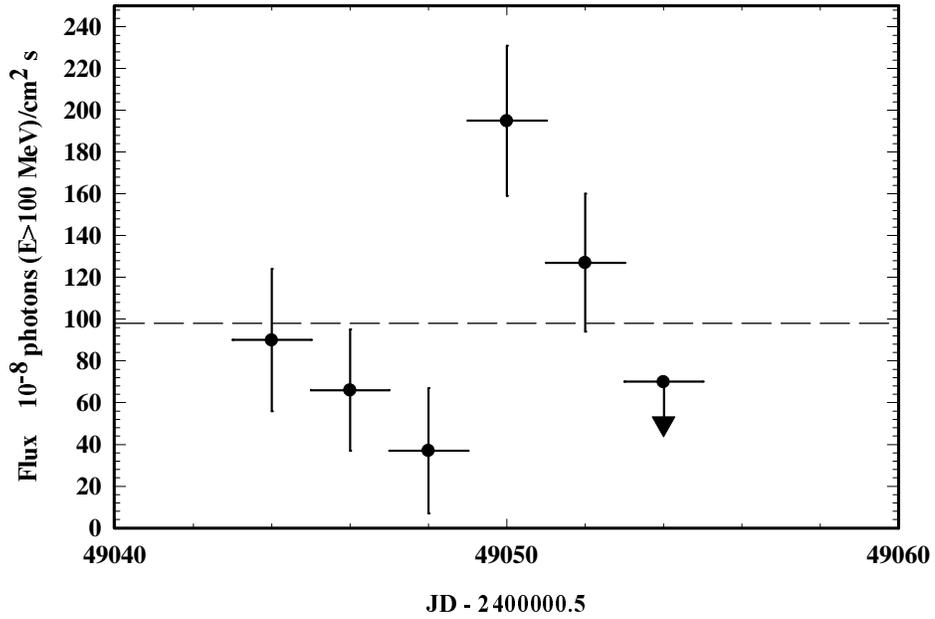}
\caption{Light curve of 3EG J0241+6103 (2CG 135+01) during VP 211.0. The points represent
         48-hour integration times. The variability index $V = 3.1$.\label{fig7}}
\end{figure}
\clearpage

\begin{figure}[t]
\epsscale{0.85}
\plotone{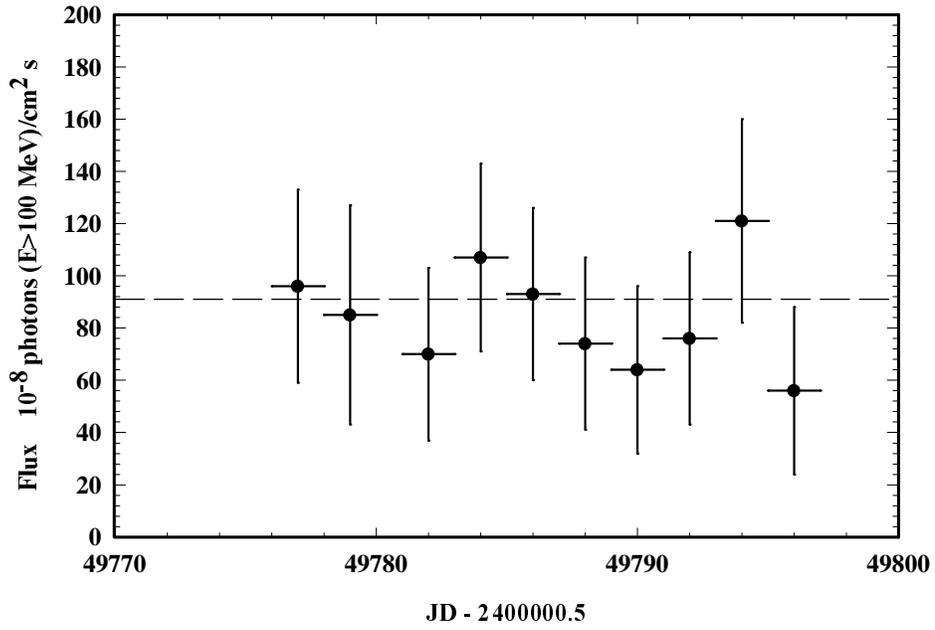}
\caption{Light curve of 3EG J0530+1323 (PKS 0528+134) during VP's 412.0 and 413.0. 
         The points represent 48-hour integration times. The variability index 
         $V = 0.013$.\label{fig8}}
\end{figure}
\clearpage

\begin{figure}[t]
\epsscale{0.85}
\plotone{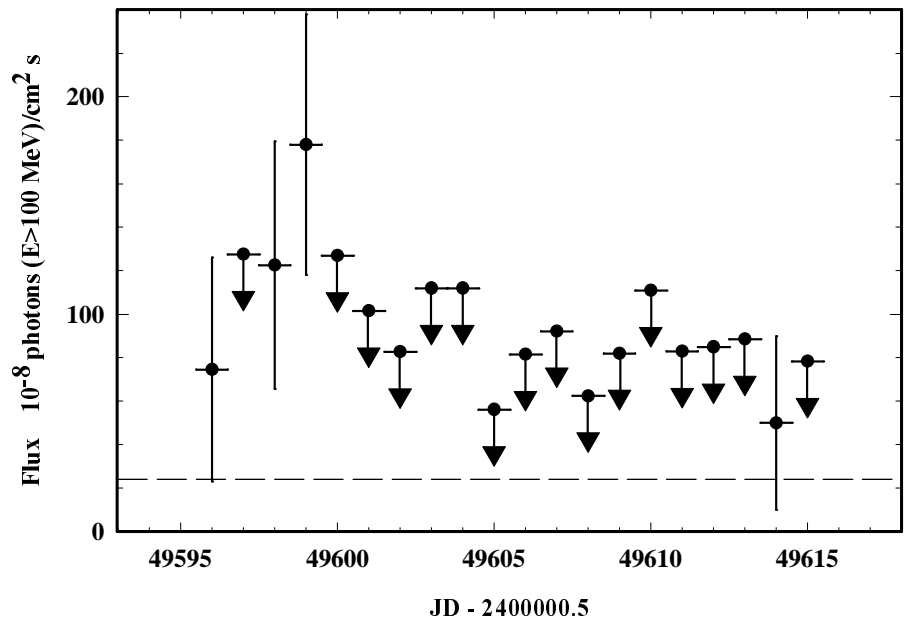}
\caption{Light curve of GRO J0927-41 during VP 338.5. The points represent
         24-hour integration times.\label{fig9}}
\end{figure}
\clearpage

\begin{figure}[t]
\epsscale{0.85}
\plotone{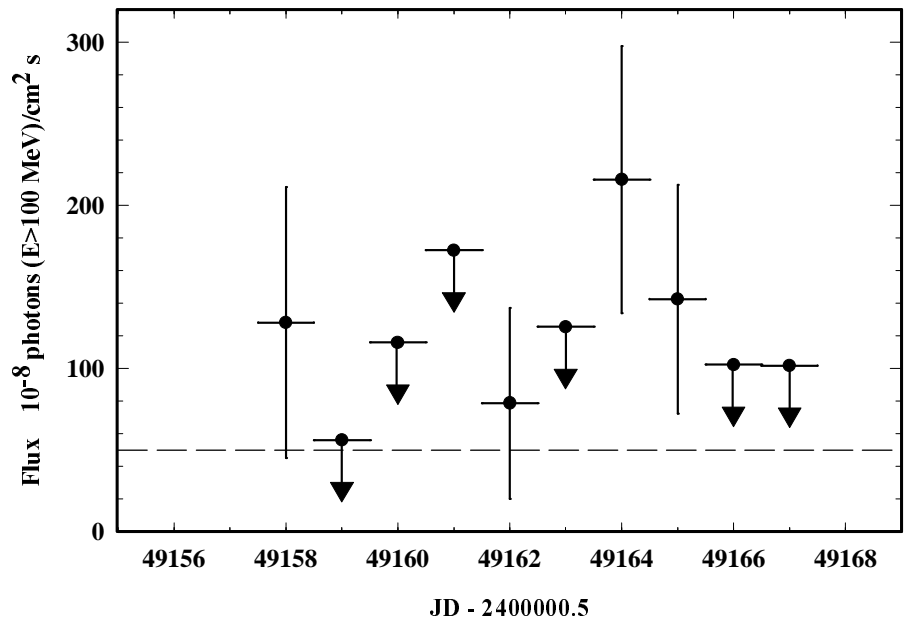}
\caption{Light curve of GRO J1547+42 during VP 226.0. The points represent
         48-hour integration times.\label{fig10}}
\end{figure}
\clearpage

\begin{figure}
\epsscale{0.8}
\plotone{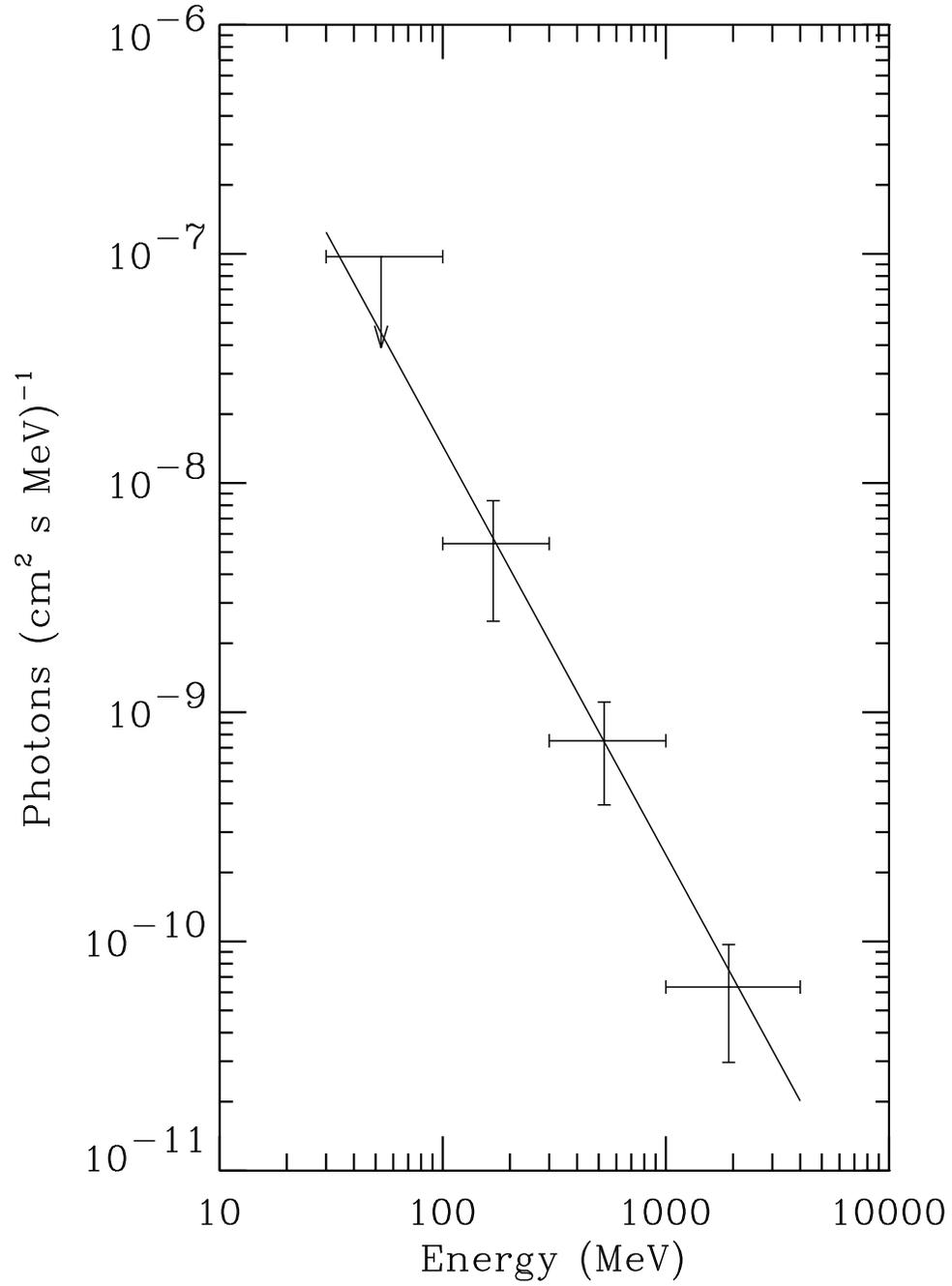}
\caption{Combined spectrum of GRO J1627-49 from VP's 23.0 and 423.5. The 
         spectral index is $1.78\pm0.31$.\label{fig11}}
\end{figure}
\clearpage

\begin{deluxetable}{ccccrcr}
\footnotesize
\tablecaption{Sources Reported in this Study. \label{tbl-1}}
\tablewidth{0pt}
\tablehead{
\colhead{Source} &\colhead{Alt Name} &\colhead{ID} &\colhead{VP's} &\colhead{$V$} 
&\colhead{Single Flare} & \colhead{$P_{mc}$}
} 
\startdata
3EG J0222+4253 &3C 66A/        &  BL Lac/&15.0       &2.6   &yes  &0.0005  \nl
               &PSR J0218+4232 &  PSR    &           &      &     &        \nl
3EG J0241+6103 &2CG 135+01     &  UID    &211.0      &3.1   &yes  &0.0006  \nl
3EG J0530+1323 &PKS 0528+134   &  FSRQ   &0.2-2.1    &4.6   &no   &n/a     \nl
GRO J0624-02\tablenotemark{a}&none&UID   &1.0,413.0  &n/a   &yes  &n/a     \nl
GRO J0927-41 &none             &  UID    &338.5      &0.5   &n/a  &0.001   \nl
3EG J1410-6147 &none           &  UID    &14.0       &2.4   &yes  &0.0006  \nl
GRO J1547-39 &none             &  UID    &226.0      &1.4   &yes  &0.010   \nl
GRO J1627-49\tablenotemark{a}&none&UID   &23.0,423.5 &n/a   &n/a  &n/a     \nl
3EG J1746-2851 &none           &  UID    &16.0       &2.2   &yes  &0.001   \nl
               &               &         &429.0      &3.0   &no   &n/a     \nl
3EG J2006-2321 &none           &  UID    &13.1       &3.2   &yes  &0.0002  \nl
\enddata

\tablenotetext{a}{Sub-threshold repeating source. Light curves were not generated
for this source since it did not display any short-duration excess greater than
4$\sigma$.}

\end{deluxetable}

\clearpage

\begin{deluxetable}{rcccc}
\footnotesize
\tablecaption{Fluxes for the Two-day Flare of 3EG J0222+4253.\label{tbl-2}}
\tablewidth{0pt}
\tablehead{
\colhead{Energy Range} &\colhead{Significance} &\colhead{Flux} &\colhead{Photons} &\colhead{Expected Photons}
} 
\startdata
100-300 MeV    &3.2$\sigma$  &$35\pm14$  &$18\pm7$    &$6\pm2$     \nl
300-1000 MeV   &4.4$\sigma$  &$22\pm8$   &$13\pm5$    &$3\pm1$     \nl
$>$1000 MeV    &1.8$\sigma$  &$4\pm3$    &$1.5\pm1.4$ &$0.2\pm0.2$ \nl
\enddata
\end{deluxetable}


\clearpage
 

\begin{thebibliography}{}
\bibitem[Bloom et al. 1997]{Blo97} Bloom, S. D. et al. 1997, Proc. 4th Comp.
         Symp., ed. C. D. Dermer, M. S. Strickman, \& J. D. Kurfess, AIP Conf.
         Proc. 410, 1262
\bibitem[Condon et al. 1998]{Con98} Condon, J. J. et al. 1998, \aj, 115, 1693
\bibitem[Douglas, J. N. et al. 1996]{Dou96} Douglas, J. N. et al. 1996, \aj, 111, 1945
\bibitem[Esposito et al. 1999]{Esp99} Esposito, J. A. et al. 1999, \apjs, 123, 203
\bibitem[Griffith, M. R. \& Wright, A. E. 1993]{Gri93} Griffith, M. R. \& Wright, A. E. 1993, \aj, 105, 1666
\bibitem[Hallum et al. 1997]{Hal97} Hallum, J. C. et al. \baas, 191, 103.15 
\bibitem[Hallum et al. 2000]{Hal00} Hallum, J. C. et al. \apj, submitted 
\bibitem[Hartman et al. 1993]{Har93} Hartman, R. C. et al. 1993, \apj, 461, 698
\bibitem[Hartman et al. 1999]{Har99} Hartman, R. C. et al. 1999, \apjs, 123, 79
\bibitem[Hartman et al. 2000]{Har00} Hartman, R. C. et al. 2000, \apj, submitted
\bibitem[Hermsen et al. 1999]{Her99} Hermsen, W. et al. Proc. 5th Comp.
         Symp., AIP Conf. Proc., in press.
\bibitem[Hunter et al. 1993]{Hun93} Hunter, S. D. et al. 1993, \apj, 409, 134
\bibitem[Hunter et al. 1997]{Hun97} Hunter, S. D. et al. 1997, \apj, 481, 205
\bibitem[Kanbach et al. 1995]{Kan95} Kanbach, G. et al. 1995, IAU Circular 6182
\bibitem[Kniffen et al. 1973]{Kni93} Kniffen, D. A. et al. 1993, \apj, 411, 133
\bibitem[Kniffen et al. 1997]{Kni97} Kniffen, D. A. et al. 1997, \apj, 486, 126
\bibitem[Mahadevan et al. 1997]{Mah97} Mahadevan, R. et al. 1997, \apj, 486, 268
\bibitem[Mattox et al. 1993]{Mat93} Mattox, J. R. et al. 1993, ApJ, 410, 609
\bibitem[Mattox et al. 1995]{Mat95} Mattox, J. R. et al. 1995, IAU Circular 6161
\bibitem[Mattox et al. 1996]{Mat96} Mattox, J. R. et al. 1996, \apj, 461, 396
\bibitem[Mattox et al. 1997a]{Mat97a} Mattox, J. R. et al. 1997a, \apj, 476, 692
\bibitem[Mattox et al. 1997b]{Mat97b} Mattox, J. R. et al. 1997b, \apj, 481, 95
\bibitem[Mattox et al. 2000]{Mat00a} Mattox, J. R. et al. 2000a, \apj, submitted
\bibitem[Mattox et al. 2000]{Mat00b} Mattox, J. R. et al. 2000b, \apjs, submitted
\bibitem[Mayer-Hasselwander et al. 1998]{May98} Mayer-Hasselwander, H. A. et al. 
         1998, A\&A 335, 161
\bibitem[McLaughlin et al. 1996]{McL96} McLaughlin, M. A. et al. 1996, \apj, 473, 763
\bibitem[Mukherjee et al. 1996]{Muk96} Mukherjee, R. et al. 1996, \apj, 470, 831
\bibitem[Mukherjee et al. 1997]{Muk97} Mukherjee, R. et al. 1997, \apj, 490, 116
\bibitem[Ramanamurthy et al. 1997]{Ram95} Ramanamurthy, P. V. et al. 1995, \apj, 450, 791
\bibitem[Sreekumar et al. 1996]{Sre96} Sreekumar, P. et al. 1996, \apjs, 464,628 
\bibitem[Sreekumar et al. 1998]{Sre98} Sreekumar, P. et al. 1998, \apj, 494, 523
\bibitem[Sturner, S. J. \& Dermer, C. D.\ 1995]{Stu95} Sturner, S. J. \& Dermer, C. D.
         1995, A\&A, 293, L17
\bibitem[Tavani et al. 1997]{Tav97} Tavani, M. et al. 1997, \apjl, 479, L109
\bibitem[Tavani et al. 1998]{Tav98} Tavani, M. et al. 1998, \apjl, 497, L89
\bibitem[Thompson et al. 1995]{Tho95} Thompson, D. J. et al. 1995, \apjs, 101, 259
\bibitem[Thompson et al. 1997]{Tho97} Thompson, D. J. et al. 1997, Proc. 4th Comp.
         Symp., ed. C. D. Dermer, M. S. Strickman, \& J. D. Kurfess, AIP Conf.
         Proc. 410, 1257
\bibitem[Verbunt et al. 1996]{Ver96} Verbunt, F. et al. 1996, A\&A, 311, L9
\bibitem[von Montigney et al. 1995]{von95} von Montigney, C. et al. 1995, \apj, 440, 525
\bibitem[Wagner et al. 1995]{Wag95} Wagner, S. et al. 1995, \apj, 454, L97
\bibitem[Wehrle et al. 1998]{Weh98} Wherle, A. et al. 1998, \apj, 497, 178 

\end{thebibliography}
\end{document}